\documentclass[
  journal=largetwo,
  manuscript=article-type,
  year=2024,
  volume=37,
]{cup-journal}

\usepackage{amsfonts}
\usepackage{amssymb}
\usepackage[nopatch]{microtype}
\usepackage[table]{xcolor}
\usepackage{booktabs}
\usepackage{physics}
\usepackage{caption}
\usepackage{hyperref}
\usepackage[natbib=true]{biblatex}
\usepackage{xcolor}
\usepackage{soul}
\usepackage{graphics}
\usepackage{float}
\usepackage{amsmath}
\usepackage{pdfpages}
\usepackage{aas-macros}
\usepackage{orcidlink}

\hypersetup{
	colorlinks=true,
	linkcolor=blue,
	filecolor=blue,
	urlcolor=blue,
	citecolor=blue
}

\title{The Three Hundred Project: The relationship between the shock and splashback radii of simulated galaxy clusters}

\author{M. Zhang\textsuperscript{\href{https://orcid.org/0000-0002-1639-7618}{\orcidlink{0000-0002-1639-7618}}}}
\affiliation{Key Laboratory of Cosmology and Astrophysics (Liaoning) College of Sciences, Northeastern University, Shenyang 110819, China}
\alsoaffiliation{International Centre for Radio Astronomy Research, The University of Western Australia, 35 Stirling Highway, Crawley, Western Australia, 6009, Australia}

\author{K. Walker\textsuperscript{\href{https://orcid.org/0000-0003-2128-1289}{\orcidlink{0000-0003-2128-1289}}}}
\affiliation{International Centre for Radio Astronomy Research, The University of Western Australia, 35 Stirling Highway, Crawley, Western Australia, 6009, Australia}
\alsoaffiliation{ARC Centre of Excellence for All Sky Astrophysics in 3 Dimensions (ASTRO 3D)}

\author{A. Sullivan\textsuperscript{\href{https://orcid.org/0009-0001-0316-0304}{\orcidlink{0009-0001-0316-0304}}}}
\affiliation{International Centre for Radio Astronomy Research, The University of Western Australia, 35 Stirling Highway, Crawley, Western Australia, 6009, Australia}
\alsoaffiliation{ARC Centre of Excellence for All Sky Astrophysics in 3 Dimensions (ASTRO 3D)}

\author{C. Power\textsuperscript{\href{https://orcid.org/0000-0002-4003-0904}{\orcidlink{0000-0002-4003-0904}}}}
\affiliation{International Centre for Radio Astronomy Research, The University of Western Australia, 35 Stirling Highway, Crawley, Western Australia, 6009, Australia}
\alsoaffiliation{ARC Centre of Excellence for All Sky Astrophysics in 3 Dimensions (ASTRO 3D)}

\author{W. Cui\textsuperscript{\href{https://orcid.org/0000-0002-2113-4863}{\orcidlink{0000-0002-2113-4863}}}}
\affiliation{Departamento de Física Teórica, Módulo 15, Facultad de Ciencias, Universidad Autónoma de Madrid, 28049 Madrid, Spain}
\alsoaffiliation{Centro de Investigación Avanzada en Física Fundamental (CIAFF), Universidad Autónoma de Madrid, Cantoblanco E-28049, Madrid, Spain }

\author{Y. Li\textsuperscript{\href{https://orcid.org/0000-0003-1962-2013}{\orcidlink{0000-0003-1962-2013}}}}
\affiliation{Key Laboratory of Cosmology and Astrophysics (Liaoning) College of Sciences, Northeastern University, Shenyang 110819, China}

\author{X. Zhang\textsuperscript{\href{https://orcid.org/0000-0002-6029-1933}{\orcidlink{0000-0002-6029-1933}}}}
\affiliation{Key Laboratory of Cosmology and Astrophysics (Liaoning) College of Sciences, Northeastern University, Shenyang 110819, China}
\alsoaffiliation{Key Laboratory of Data Analytics and Optimization for Smart Industry (Ministry of Education), Northeastern University, Shenyang 110819, China}
\alsoaffiliation{National Frontiers Science Center for Industrial Intelligence and Systems Optimization, Northeastern University, Shenyang 110819, China}

\email[C. Power and X. Zhang]{chris.power@uwa.edu.au; zhangxin@mail.neu.edu.cn}


\begin{document}

\begin{abstract}
Observations of the intracluster medium (ICM) in the outskirts of galaxy clusters reveal shocks associated with gas accretion from the cosmic web. Previous work based on non-radiative cosmological hydrodynamical simulations have defined the shock radius, $r_\text{shock}$, using the ICM entropy, $K \propto T/{n_\mathrm{e}}^{2/3}$, where $T$ and $n_\text{e}$ are the ICM temperature and electron density respectively; the $r_\text{shock}$ is identified with either the radius at which $K$ is a maximum or at which its logarithmic slope is a minimum. We investigate the relationship between $r_\text{shock}$, which is driven by gravitational hydrodynamics and shocks, and the splashback radius, $r_\text{splash}$, which is driven by the gravitational dynamics of cluster stars and dark matter and is measured from their mass profile. Using 324 clusters from {\small The Three Hundred} project of cosmological galaxy formation simulations, we quantify statistically how $r_\text{shock}$ relates to $r_\text{splash}$. Depending on our definition, we find that the median $r_\text{shock} \simeq 1.38 r_\text{splash} (2.58 R_{200})$ when $K$ reaches its maximum and $r_\text{shock} \simeq 1.91 r_\text{splash} (3.54 R_{200})$ when its logarithmic slope is a minimum; the best-fit linear relation increases as $r_\text{shock} \propto 0.65 r_\text{splash}$. We find that $r_\text{shock}/R_{200}$ and $r_\text{splash}/R_{200}$ anti-correlate with virial mass, $M_{200}$, and recent mass accretion history, and $r_\text{shock}/r_\text{splash}$ tends to be larger for clusters with higher recent accretion rates. We discuss prospects for measuring $r_\text{shock}$ observationally and how the relationship between $r_\text{shock}$ and $r_\text{splash}$ can be used to improve constraints from radio, X-ray, and thermal Sunyaev-Zeldovich surveys that target the interface between the cosmic web and clusters.
\end{abstract}

\section{Introduction}
Galaxy clusters are the most massive virialised structures in the present-day Universe; in hierarchical cosmologies such as the $\Lambda$ Cold Dark Matter model they assemble relatively recently, with typical formation redshifts of $z_\text{form}\simeq 0.5$ \citep[e.g.][]{2007MNRAS.375..489H,2008MNRAS.389.1419L,Power.etal.2012}. Clusters sit at the nodes of the cosmic web, accreting material from filaments, which is evident in the relative positions and orbits of infalling galaxies and groups \citep[e.g.][]{Tempel.etal.2015} and in accretion shocks in the hot intracluster medium \autocite[hereafter ICM; e.g.][]{Burns.etal.2010,Brown.Rudnick.2011,Power.etal.2020}.

A commonly used measure of the physical state of a cluster's ICM is the entropy, $K$, which is defined as,
\begin{equation}
    K \equiv \frac{k_\mathrm{B}T}{n_\mathrm{e}^{2/3}}
    \label{entropy}
\end{equation}
\citep[cf.][]{Cavaliere.Lapi.2013}; here $k_\mathrm{B}$ is the Boltzmann constant, $T$ is the ICM gas temperature, and $n_\mathrm{e}$ is the electron number density, which is related to the ICM gas density. High-resolution X-ray observations, including XMM-\textit{Newton} \autocite[e.g.][]{XMMNewton}, \textit{Chandra} \autocite[e.g.][]{Chandra}, and eROSITA \autocite[e.g.][]{eROSITA} have allowed the radial variation of cluster entropy to be studied in detail, and consequently the functional form of the entropy with respect to the radius $r$, $K(r)$, is well-understood \citep[e.g.][]{Panagoulia.etal.2014,Hogan.etal.2017,McDonald.etal.2019,Zhu.etal.2021}. $K(r)$ can be characterised by its logarithmic slope, $k$, which is defined as,
\begin{equation}
    k \equiv \frac{\mathrm{d} \ln K}{\mathrm{d} \ln r},
\end{equation}
which is itself a function of radius. 

Observationally we find that $K$ is consistent with being a power-law near $R_{500}$ such that $k \simeq 1.1$ \autocite[see, e.g.][]{Babyk2018, Ghirardini2019}; here $R_\text{500,crit}$ is the radius at which the enclosed matter density is 500 times the critical density, $\rho_\text{crit}=3H^2/8\pi\,G$. This power-law behaviour is recovered in hydrodynamical cosmological simulations \citep[e.g.][]{Voit2005}, independent of hydrodynamics solver and galaxy formation model \citep[e.g.][]{Sembolini.etal.2016b}. At larger radius, simulations predict that $K(r)$ reaches a maximum at $\simeq\,1.6\,R_\text{200,mean}$, where $R_\text{200,mean}$ encloses a mean matter density that is 200 times the cosmological mean matter density, $\rho_\text{mean}=\Omega_\text{m}\rho_\text{crit}$, where $\Omega_\text{m}$ is the matter density parameter \autocite[cf.][]{2015ApJ...806...68L}. We note that this predicted radius is larger than that inferred from observational data \citep[cf.][]{Walker.etal.2012}, which indicate that the entropy profile reaches its maximum closer to $R_\text{200,crit}\equiv\,R_{200}$, the radius enclosing a mean matter density of 200$\rho_\text{crit}$. Regardless, the presence of a turnover in the entropy profile is interpreted as arising from infalling gas from the cosmic web generating an accretion shock at the `shock radius', $r_\text{shock}$, which is consistent with empirical measurements of the interface between cluster outskirts and filaments in the cosmic web \citep[e.g.][]{Kawaharada.etal.2010}. For this reason we can regard $r_\text{shock}$ as a characteristic measure of the boundary between a cluster's accreted gas reservoir and gas in the process of accreting from the cosmic web. 

The splashback radius, $r_\text{splash}$, provides an analogous characteristic measure of the boundary between collisionless material - dark matter and galaxies - that is orbiting within a cluster's potential and material that is infalling for the first time \citep[e.g.][]{More.etal.2015,Mansfield.etal.2017,Diemer.etal.2017,Deason.etal.2021}. 
By convention, $r_\text{splash}$ is defined as the radius at which the logarithmic slope of the spherically averaged density profile reaches its minimum value. Observational estimates of $r_\text{splash}$ using, for example, the luminosity density profile, galaxy number densities, and weak lensing measurements \citep[e.g.][]{Chang.etal.2018,Bianconi.etal.2021,Gonzalez.etal.2021} indicate good consistency between simulation predictions and observationally inferred values, although observational estimates will be sensitive to a cluster's dynamical state and the structure of the cosmic web in which it is embedded \citep[e.g.][]{Lebeau.etal.2024}.

The question arises naturally as to the relationship between $r_\text{shock}$ and $r_\text{splash}$. Both are characteristic of the growth of clusters by the accretion of material from their surroundings. Analytical models have assumed that $r_\text{shock}$ and $r_\text{splash}$ are coincident \autocite[e.g.][who assume that the shock in the hot gas profile is coincident with a break in the dark matter profile]{Patej&Loeb2015}. However, $r_\text{shock}$ arises because of the collisional nature of accreting gas whereas $r_\text{splash}$ is a result of the complex dynamics of collisionless components in an evolving gravitational potential, and so it's likely that instances in which $r_\text{shock}$ and $r_\text{splash}$ are coincident and infrequent at cluster mass scales. 

The goal of this paper is to quantify the relationship between $r_\text{shock}$ and $r_\text{splash}$ and its predicted variation with mass and recent accretion history using a statistical sample of massive galaxy clusters from {\small The Three Hundred} collaboration's simulation suite \citep[cf.][]{Cui.etal.2018,Cui.etal.2022}. This is a mass complete sample of clusters drawn from a 1 $h^{-1}$ Gpc box, which have a diversity of assembly histories and larger-scale environments. 

In the following sections, we describe briefly {\small The Three Hundred} project and our approach to calculating $r_\text{shock}$ and $r_\text{splash}$ (\S~\ref{sim}). We present the measured relationship between $r_\text{shock}$ and $r_\text{splash}$ and their variation with cluster mass and accretion rate (\S~ \ref{results}), and we discuss our results in the context of previous work (\S~\ref{discussion}). Finally, we summarise our main findings in \S~\ref{conclusion}.

\section{The Simulated Dataset}\label{sim}
We use the 324 clusters from the latest {\small GIZMO-Simba} runs - hereafter {\small GIZMO-Simba-7k} (Cui et al., In Preparation) - of {\small The Three Hundred} collaboration's suite of zoom simulations of galaxy clusters \citep[cf.][]{Cui.etal.2018}. These are a higher resolution extension - with re-calibrated galaxy formation prescriptions - of the {\small GIZMO-Simba} runs - hereafter {\small GIZMO-Simba-3k} - presented in \citet{Cui.etal.2022}. {\small GIZMO-Simba-3k} modelled galaxy formation processes (radiative cooling, star formation and feedback, black hole formation and growth, multiple modes of black hole feedback) using a variant of the {\small SIMBA} galaxy formation model presented in \citet{Dave.etal.2019}, calibrated for cluster scales as detailed in \citet{Cui.etal.2022}, and run with {\small GIZMO} \citep{Hopkins.2015}. {\small GIZMO-Simba-7k} uses an updated version of the {\small SIMBA} model -- {\small SIMBA-C} \citep{Hough2023}, which adopts the advanced chemical enrichment model of \citet{Kobayashi2020}. {\small SIMBA-C} also includes several other modifications, including a jet velocity that depends on the host dark matter halo's mass via the approximate escape velocity and a lower black hole seeding mass ($M_\ast \gtrsim 6 \times 10^6 M_\odot$ compared to $M_\ast \gtrsim 5 \times 10^9 M_\odot$ in {\small SIMBA}). We refer interested readers to \citet{Hough2023} for more details. We note that the calibration of {\small GIZMO-Simba-7k} considered both the stellar and gas properties of the cluster, unlike {\small GIZMO-Simba-3k}, which was calibrated against only stellar properties; this produce improved ICM properties in {\small GIZMO-Simba-7k}.

These clusters form a mass complete sample at $z$=0 in the MultiDark Planck 2 simulation \citep[][]{Klypin.etal.2016}, a $1 h^{-1}$ Gpc box on a side. They have virial masses in the range $6.4 \times 10^{14} h^{-1}M_{\odot} \lesssim M_{200} \lesssim 2.6 \times 10^{15} h^{-1}M_{\odot}$, where $M_{200}$ is the mass corresponding to an overdensity criterion of 200 times the critical density at that epoch. The zoom region extends 15 $h^{-1}$Mpc from the centre of the cluster at $z$=0, corresponding to several virial radii; dark matter and gas cell masses in this region are $m_{\rm dm} \simeq 10^8 h^{-1}M_{\odot}$ and $m_{\rm gas} \simeq 2 \times 10^7 h^{-1}M_{\odot}$ respectively. The adopted cosmological parameters are $(\Omega_\text{m},\Omega_\text{B},\Omega_{\Lambda},h,\sigma_8)=(0.307, 0.048, 0.693, 0.678, 0.823)$.

\medskip
For each cluster, we use group catalogues constructed with the {\small AHF} halo finder \citep[cf.][]{Knollman.Knebe.2009}, which includes information about the stellar and gas content of the main halo and its substructures. We compute radial profiles for the mass density and gas entropy using 100 equally spaced logarithmic bins between minimum and maximum cluster-centric radii of 0.5$R_{200}$ and 5$R_{200}$; here the centre is the density-weighted centre of the adaptive mesh refinement grid in {\small AHF}. The presence of substructure in the outskirts of the halo can bias estimates of the logarithmic slope, which influences the value of $r_\text{splash}$. To avoid this, we evaluate the density and mass-weighted temperature in 48 angular segments within each radial bin and take the median value within the bin as our estimate of the density and temperature at that radius. With this information we can estimate the entropy in a given radial bin following Equation \ref{entropy}. Note that for each gas element\footnote{A gas element can refer to either a gas particle, as in the {\small GadgetX} model (see Appendix), or a gas cell, as in the {\small Gizmo-Simba} models.} in the cluster we have an associated internal energy per unit mass, $u=3k_{\rm B}T/(2\mu m_{\rm p})$, from which we can estimate the temperature, and a local density, $\rho=\mu_{\rm e}m_{\rm p}n_{\rm e}$; here $\mu$ and $\mu_{\rm e}$ are the mean molecular weights of the gas and the electrons, respectively, and $m_{\rm p}$ is the proton mass. The profiles and their logarithmic slopes are smoothed by a Gaussian filter to allow for reliable identification of maxima and minima.

Note that there are two different definitions for shock radius in the literature - one defined by the radius at which the entropy profile reaches its maximum or "peak" \citep[cf.][]{2015ApJ...806...68L}, which we indicate by $r_\text{shock,p}$, and one defined by the radius at which the logarithmic slope of the entropy profile is a minimum \citep[cf.][]{2016MNRAS.461.1804S}, which we indicate by $r_\text{shock,m}$. We provide predictions for both $r_\text{shock,p}$ and $r_\text{shock,m}$.

\section{Results}\label{results}
\begin{figure}[tp]
\centering
\includegraphics[width=0.9\columnwidth]{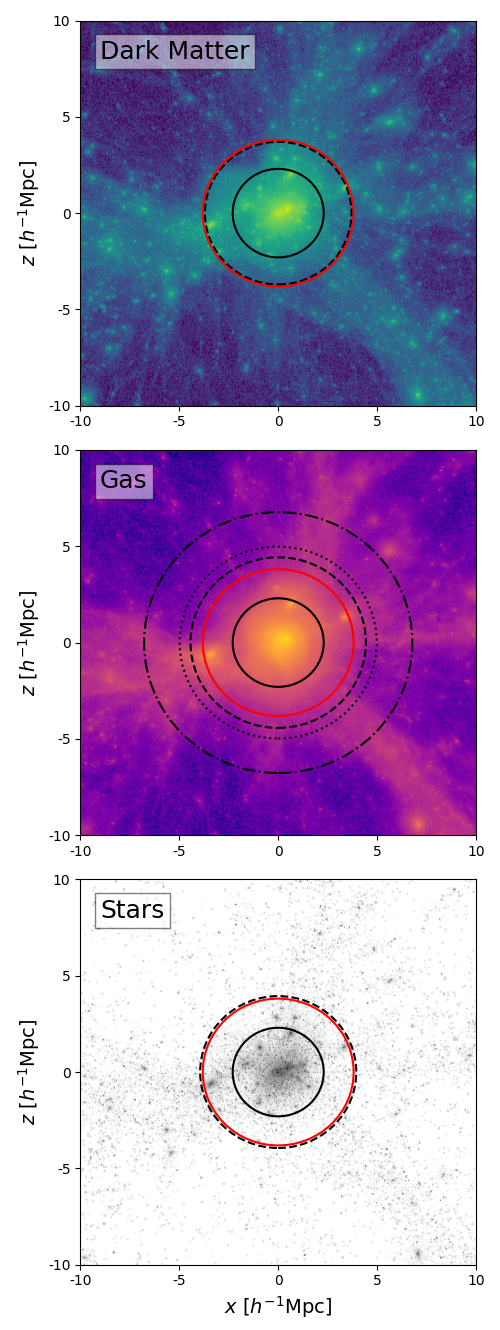}
\caption{Projected dark matter, gas, and stellar densities (top to bottom) at $z$=0 in the most massive cluster in our sample within a cubic region 20 $h^{-1}$Mpc, centred on the density-weighted centred of {\small AHF}'s adaptive mesh refinement grid. The dark matter halo's mass and radius are $M_{200}=2.82\times 10^{15} h^{-1}M_{\odot}$ and $R_{200, crit}=2.298 h^{-1} \rm Mpc$, and it has accreted 75\% of its present day mass since $z$=0.5.} Heavy solid, red solid, dashed, dotted and dot-dashed circles indicate $R_{\rm 200,crit}$, $R_{\rm 200,mean}$, $r_\text{splash}$, $r_\text{shock,p}$, and $r_\text{shock,m}$ respectively.
\label{figure:visual_impression_cluster1}
\end{figure}

\begin{figure}[h]
\centering
\includegraphics[width=0.9\columnwidth]{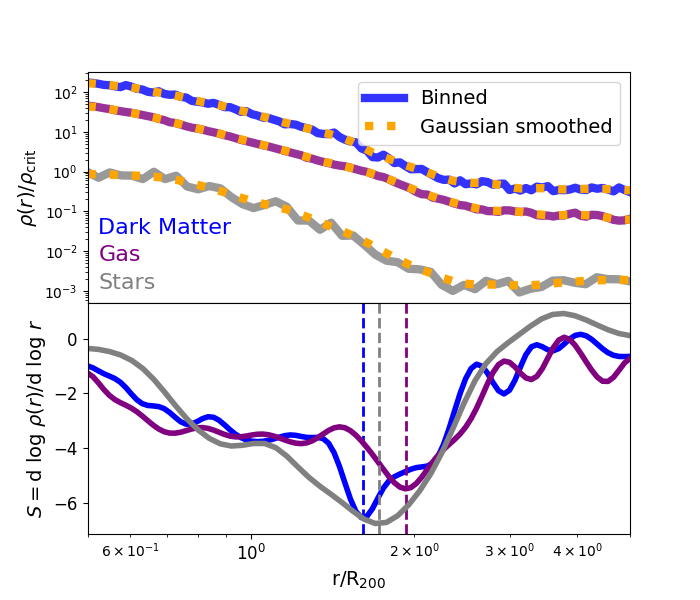}
\includegraphics[width=0.9\columnwidth]{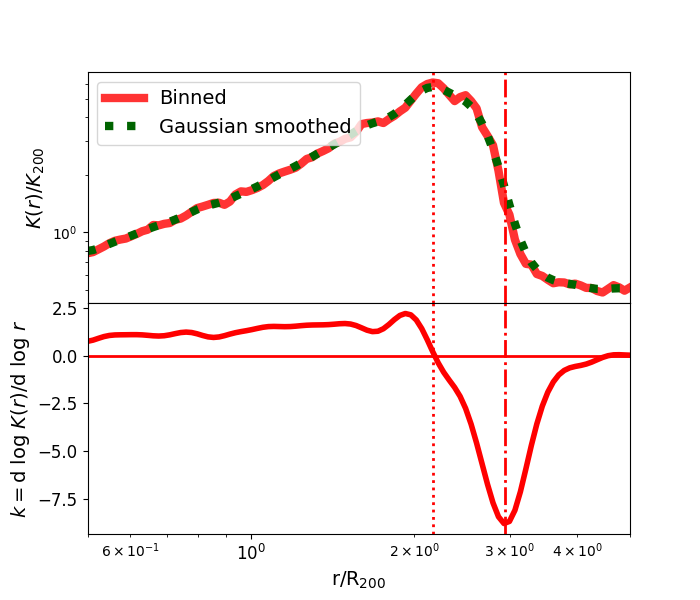}
\caption{Density (top) and gas entropy (bottom) radial profiles, along with their logarithmic slopes (lower panels) for the cluster shown in Figure~\ref{figure:visual_impression_cluster1}. Dashed vertical lines in the top panel correspond to $r_\mathrm{splash}=1.61 R_{200}$ (blue) for dark matter, $r_\mathrm{splash}=1.93 R_{200}$ (purple) for gas, $r_\mathrm{splash}=1.71 R_{200}$ (gray) for stars, respectively. Dotted, and dot-dashed vertical lines in the bottom panel correspond to $r_\text{shock,p}=2.17 R_{200}$ and $r_\text{shock,m}=2.95 R_{200}$ respectively.}
\label{figure:profiles_cluster1}
\end{figure}

We begin with a visual impression of the most massive cluster in our sample, showing the relative positions of $r_\text{splash}$, $r_\text{shock,p}$, and $r_\text{shock,m}$. This cluster has a $z$=0 virial mass of $M_{200}$=$2.82\times 10^{15} h^{-1}M_{\odot}$ and virial radius of $R_{200, \rm crit}=2.298 h^{-1} \rm Mpc$. Although it is not currently undergoing a significant merger, it has accreted 75\% of its mass since $z$=0.5, which indicates that it has a high recent accretion rate. In Figure~\ref{figure:visual_impression_cluster1}, we show  projections of the distribution of dark matter (top panel), gas (middle panel), and stars (lower panel) around the most massive cluster in our sample at $z$=0 within a comoving cube of size 20 $h^{-1}$Mpc. In each of the panels, the heavy solid, red solid, and heavy dashed circles indicate the virial radii, $R_{200}=2.30 h^{-1} \rm{Mpc}$ and $R_{\rm 200,mean}=3.81 h^{-1} \rm{Mpc}$, and splashback radius, $r_\text{splash}=1.61 R_{200}$ for dark matter; if not specified, $R_{\rm 200}$ indicates $R_{\rm 200,crit}$ throughout this paper. The heavy dotted and dot-dashed circle in the middle panel (projected gas distribution) indicate the two definitions of shock radius, $r_\text{shock,p}=2.17 R_{200}$ and $r_\text{shock,m}=2.95 R_{200}$, respectively. For completeness, we also estimate $r_\text{splash}=1.93 R_{200}$ for the gas profile and $r_\text{splash}=1.71 R_{200}$ for stellar profile.  It's interesting to note that, for this particular cluster, the splashback radius of the gas is within $\sim$10\% of the shock radius defined relative to the peak of the entropy profile; the splashback radii of the dark matter and stars are within 5\% of each other, as we might expect given their collisionless nature; and the splashback radius of the dark matter is $\sim 20\%$ smaller than that of the gas. For reference, the radial (dark matter, stellar, gas) density and gas entropy profiles used to estimate $r_\text{splash}$ and $r_\text{shock}$ are shown in Figure~\ref{figure:profiles_cluster1}.

\begin{figure}[h]
\centering
\includegraphics[width=0.9\columnwidth]{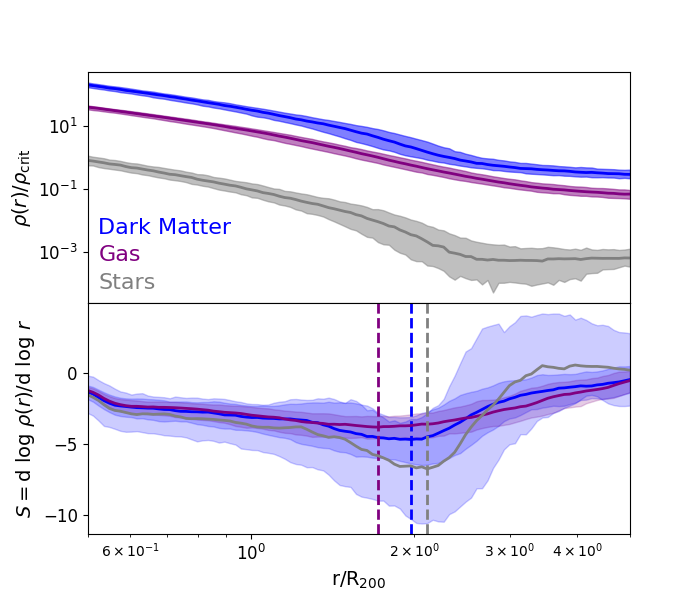}
\includegraphics[width=0.9\columnwidth]{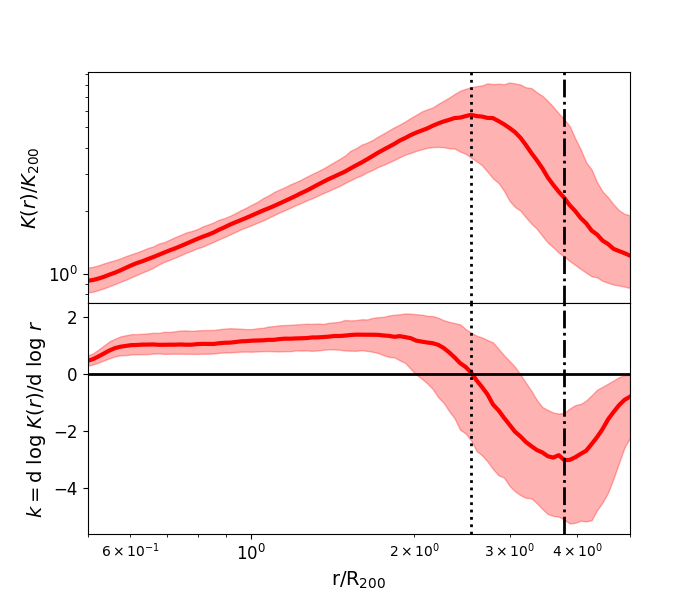}
\caption{Radial profiles of dark matter density (top) and gas entropy (bottom) with their logarithmic slopes for all {\small The Three Hundred} collaboration's suite of simulated clusters. The curves and shaded regions correspond to the median and the range between the $10^{\rm th}$ to $90^{\rm th}$ percentiles from the distribution of cluster profiles. The dashed line in the top panel represents the location of $r_\text{splash}$. The dotted and dot-dashed lines in the bottom panel indicate the location of $r_\text{shock,p}$ and $r_\text{shock,m}$, respectively. Curves are colour coded as in Figure~\ref{figure:profiles_cluster1}.}
\label{figure:300profiles}
\end{figure}

\begin{figure*}[ht]
\centering
\includegraphics[width=0.95\linewidth]{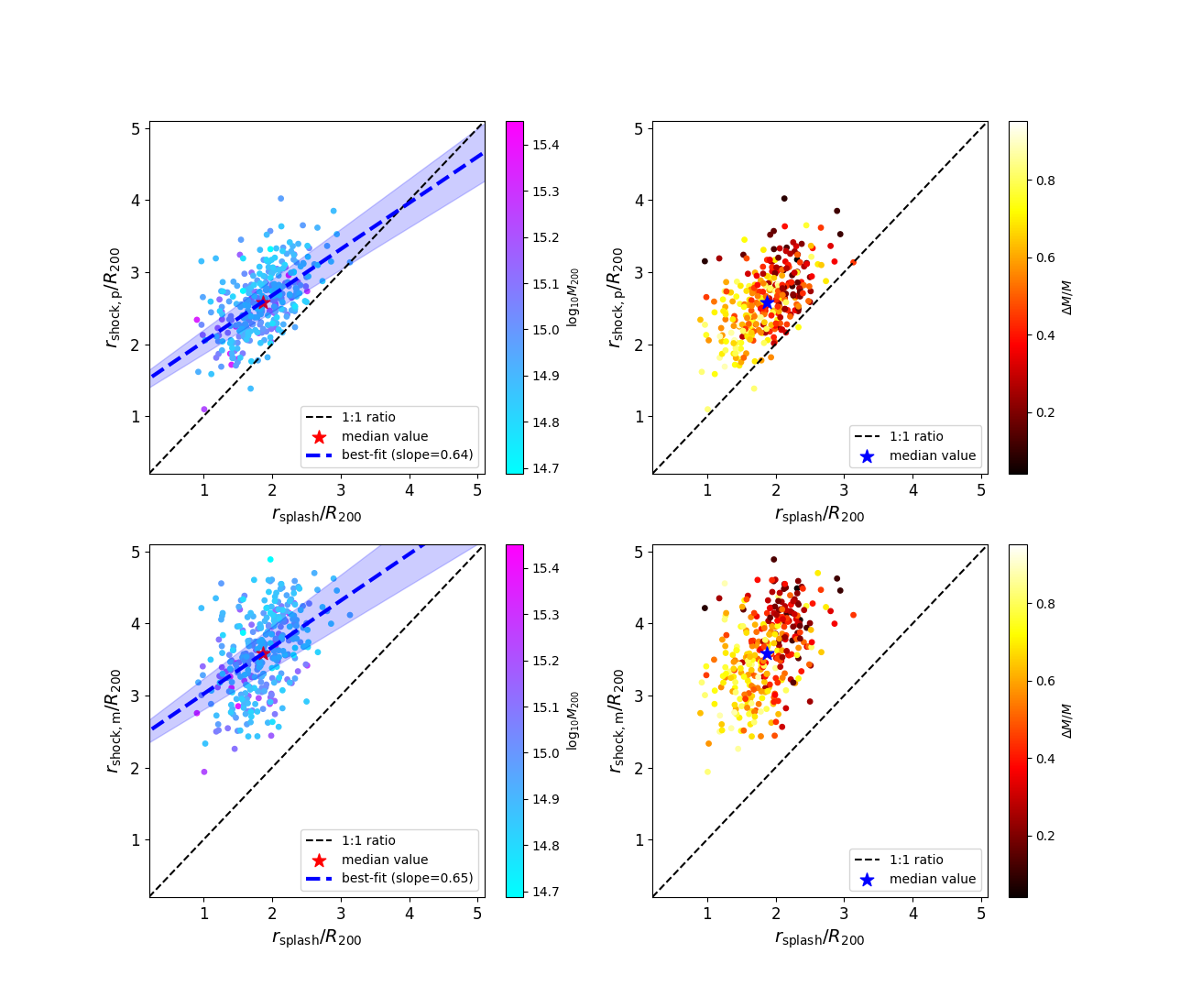}
\caption{The relationship between the shock radius $r_\text{shock}$ and splashback radius $r_\text{splash}$ for each of the 324 clusters in our sample. Upper panels correspond to $r_\text{shock,p}$ identified with the maximum of $K$, while the lower panels correspond to $r_\text{shock,m}$ identified with the minimum of its logarithmic slope. The points are colour coded by the virial mass $M_{200}$ (left panels) and the fractional increase in $M_{200}$ since $z=0.5$, $\Delta M / M$ (right panels). The red and blue stars indicate the median values $r_\text{shock}$ and $r_\text{splash}$, while the light and heavy dashed lines correspond to the one-to-one relationships and the best-fit linear relationships. The shaded band in the left-hand panels indicates the 1-$\sigma$ variation estimated by bootstrapping.}
\label{figure:Rsp_Rsh}
\end{figure*}

\begin{figure}[ht!]
\centering
\includegraphics[width=0.9\columnwidth]
{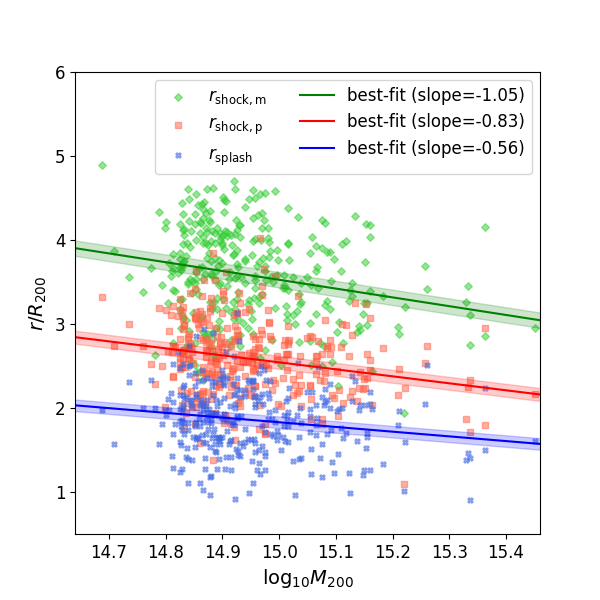}
\includegraphics[width=0.9\columnwidth]
{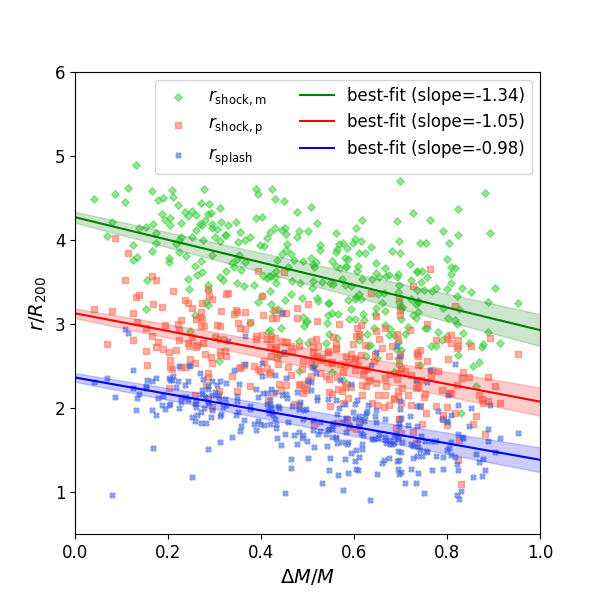}
\caption{The relationship between the shock and splashback radii, $r_\text{shock, p}$, $r_\text{shock, m}$, and $r_\text{splash}$ as a function of virial mass, $M_{200}$ (upper panel) and recent mass accretion history (lower panel) for each of the 324 clusters in our sample. The shaded bands indicate the 1-$\sigma$ variations for each set of points estimated by bootstrapping.}
\label{figure:mass_accretion}
\end{figure}

This trend - for $R_{200}<r_\text{splash}<r_\text{shock,p}<r_\text{shock,m}$ - is characteristic of the typical cluster in this sample. We demonstrate this in Figure~\ref{figure:300profiles} in which we show the median dark matter density profiles (blue curves and shaded regions; top) and gas entropy profiles (red curves and shaded regions; bottom) and their corresponding logarithmic slopes for all 324 clusters; the shaded bands indicate the range of variation between the 10$^\text{th}$ and 90$^\text{th}$ percentiles within each radial bin. The dashed, dotted, and dot-dashed lines indicate the locations of $r_\text{splash}$, $r_\text{shock,p}$, and $r_\text{shock,m}$ of the median cluster. We find that $r_\text{splash}=1.87^{+0.39}_{-0.41} R_{200}$, $r_\text{shock,p}=2.58^{+0.45}_{-0.43} R_{200}$ and $r_\text{shock,m}=3.58^{+0.57}_{-0.62} R_{200}$. We also include the splashback radii for gas - $r_\text{splash}/R_{200} = 1.72^{+0.55}_{-0.45}$ - and stars - $r_\text{splash}/R_{200} = 2.07^{+0.42}_{-0.39}$ - in Figure~\ref{figure:300profiles} (purple and grey curves and shaded regions, top). This means that the stellar mass density profile traces that of the underlying dark matter, and indeed we see a stronger splashback feature in the stars. In contrast, the gas mass density profile is smoother and shows no obvious feature, and formally reaches a minimum slope at a smaller radius compared to the stars although this is not a strong feature.

For this typical cluster, we note that the relative coincidence of $r_\text{splash}$ for the gas and $r_\text{shock,p}$ evident in Figure~\ref{figure:profiles_cluster1} is absent. This is not so surprising because we expect strong cluster-to-cluster variations in density and temperature in the outskirts of clusters \citep[e.g.][]{Power.etal.2020}, which will dampen any splashback features in the median gas density profile. We defer a more detailed study of the relationship between gas splashback and shock radii to a subsequent paper. Note that from here and for the remainder of the paper, when we refer to $r_\text{splash}$ we use the value defined for the dark matter.

The clusters in our sample have diverse assembly histories and larger-scale environments, and so we expect cluster-to-cluster variations in $r_\text{splash}$ and $r_\text{shock}$. We quantify this in Figure~\ref{figure:Rsp_Rsh} in which we show how $r_\text{shock,p}$ (upper panels) and $r_\text{shock,m}$ (lower panels) vary with $r_\text{splash}$ for each cluster, in units of $R_{200}$; on the left we investigate trends with virial mass, $M_{200}$, while on the right we look at trends with the fractional increase in $M_{200}$ since $z=0.5$, 
\begin{equation}
\frac{\Delta\,M}{M}=\frac{M_{200}(z=0)-M_{200}(z=0.5)}{M_{200}(z=0)}.
\label{eq:deltaM}
\end{equation}
Compared to $\Gamma$, the accretion rate conventionally used in the literature \citep[cf. the equation 1 of][]{2021MNRAS.506..839Z}\footnote{The conventional accretion rate is,
\begin{equation}
\Gamma=\frac{d\log\!M}{d\log\!a}\equiv\frac{\Delta\!M/M}{\Delta\log(a)}
\label{eq:gamma}
\end{equation}
where $a=1/(1+z)$ is the expansion factor. We assume that $\Delta\log(a)\equiv-\log(a)$ when considering a change in mass with respect to $z$=0, $a$=1.}, Equation~\ref{eq:deltaM} corresponds to $\Gamma\log(1+z)\simeq\!0.41\Gamma$ for $z$=0.5. 

Figure~\ref{figure:Rsp_Rsh} reveals that $r_\text{shock,p}$ is larger than $r_\text{splash}$ for all but a handful of clusters, while $r_\text{shock,m}$ is consistently larger than $r_\text{splash}$ for all cases. The stars indicate the median values of $r_\text{shock}$ and $r_\text{splash}$ for the sample are $r_\text{shock,p}/r_\text{splash}=1.38^{+0.27}_{-0.21}$ and $r_\text{shock,m}/r_\text{splash}=1.91^{+0.31}_{-0.42}$, while the best-fit linear relationships between $r_\text{shock}$ and $r_\text{splash}$ - which we show in the left hand panels - are
\begin{equation}
\label{eq:shockp_splash}
r_\text{shock,p} = 0.64(\pm0.06)\,r_\text{splash} + 1.39(\pm0.11),
\end{equation}
and
\begin{equation}
\label{eq:shockm_splash}
r_\text{shock,m} = 0.65(\pm0.07)\,r_\text{splash} + 2.38(\pm0.14),
\end{equation}
where $r_\text{splash}, r_\text{shock,p}$ and $r_\text{shock,m}$ are in units of $h^{-1}$Mpc. We estimate $1-\sigma$ uncertainties via bootstrapping; these are listed in parentheses and by the shaded bands around the best-fit lines in Figure~\ref{figure:Rsp_Rsh}. The Spearman rank correlation coefficients are $r_s$=0.469 and 0.564 (with vanishingly small $p$-values) for Equations \ref{eq:shockp_splash} and \ref{eq:shockm_splash}, respectively; this indicates that there is a moderate positive correlation between the shock radii and the splashback radius.

Figure~\ref{figure:Rsp_Rsh} shows how $r_\text{splash}$ and $r_\text{shock}$ relate to one another for a given cluster and trends between this relationship and the cluster's virial mass, $M_{200}$, and its recent fractional change in $M_{200}$, $\Delta\,M/M$. In Figure~\ref{figure:mass_accretion} we quantify the trends between $r_\text{splash}$ and $r_\text{shock}$ with $M_{200}$ and $\Delta\,M/M$ directly. 
The top panel shows how $r_\text{splash}$ and $r_\text{shock}$ vary with $M_{200}$, in units of $R_{200}$, for all 324 clusters in our sample. The data can be characterised by the relations,
\begin{equation}
\label{eq:shockp_m200}
\frac{r_\text{shock,p}}{R_{\rm 200}} = -0.82 \, \log_{10} M_{\rm 200} + 15.02(\pm0.08),
\end{equation}
\begin{equation}
\label{eq:shockm_m200}
\frac{r_\text{shock,m}}{R_{\rm 200}} = -1.05 \, \log_{10} M_{\rm 200} + 19.23(\pm0.09), 
\end{equation}
and 
\begin{equation}
\label{eq:splash_m200}
\frac{r_\text{splash}}{R_{\rm 200}} = -0.56 \, \log_{10} M_{\rm 200} + 10.23(\pm0.07),
\end{equation}
where, as before, $M_{200}$ is in units of $h^{-1}M_{\odot}$. 1-$\sigma$ uncertainties, estimated via bootstrapping, are in parentheses, and are shown as shaded bands in the Figure. We find Spearman rank correlation coefficients of $r_s=-0.20$ for $r_\text{shock,p}$, $r_s=-0.18$ for $r_\text{shock,m}$ and $r_s=-0.14$ for $r_\text{splash}$, with respect to $M_{200}$, which indicates that there is a weak anti-correlation with virial mass.

The bottom panel shows shows how $r_\text{splash}$ and $r_\text{shock}$ vary with $\Delta\,M/M$, in units of $R_{200}$. We find,
\begin{equation}
\label{eq:shockp_deltam}
\frac{r_\text{shock,p}}{R_{\rm 200}} = -1.05(\pm0.11) \, \Delta M/M + 3.13(\pm0.06),
\end{equation}
\begin{equation}
\label{eq:shockm_deltam}
\frac{r_\text{shock,m}}{R_{\rm 200}} = -1.34(\pm0.12) \, \Delta M/M + 4.27(\pm0.07),
\end{equation}
and
\begin{equation}
\label{eq:splash_deltam}
\frac{r_\text{splash}}{R_{\rm 200}} = -0.98(\pm0.10) \, \Delta M/M + 2.36(\pm0.05).
\end{equation}
As above, 1-$\sigma$ uncertainties are in parentheses and are shown as shaded bands in the Figure. The Spearman rank correlation coefficients are $r_s=-0.48$ for $r_\text{shock,p}$, $r_s=-0.51$ for $r_\text{shock,m}$ and $r_s=-0.57$ for $r_\text{splash}$, which indicate a moderate anti-correlation with our measure of the recent accretion rate.

These trends, along with the best-fit linear relationships (Equations~\ref{eq:shockp_splash} and \ref{eq:shockm_splash}), indicate that there is a moderate positive correlation between $r_\text{shock}$ and $r_\text{splash}$, driven by a cluster's recent mass accretion rate. This is consistent with the findings of \citet{2021MNRAS.508.2071A}. These findings are largely insensitive to mass resolution and galaxy formation model, provided care is taken to recover ICM properties that are consistent with observations. We discuss this in more detail in \ref{sec:appendix}.

\section{Discussion}\label{discussion}
There has been significant progress over the last decade in our understanding of the physical processes that shape the outskirts of galaxy clusters, using both cosmological simulations and a variety of observational data. Radio synchrotron emission and polarisation \citep[e.g.][]{Locatelli.etal.2021,Ha.etal.2023,Vernstrom.etal.2023,Boss.etal.2023}, gas entropy \citep{2015ApJ...806...68L,2021MNRAS.508.2071A} and X-ray emission \citep[e.g.][]{Simionescu.etal.2021}, and the thermal Sunyaev–Zeldovich (tSZ) effect \citep{2021MNRAS.508.1777B,2022MNRAS.514.1645A,2024MNRAS.527.9378A} all offer the means to probe the shocked gas associated with accretion from the cosmic web. That there is a relationship between this accretion shock and the cluster boundary defined by the splashback radius has been explored observationally \citep{2022MNRAS.514.1645A,2024MNRAS.527.9378A} and in non-radiative cosmological hydrodynamical simulations \citep{2019SSRv..215....7W,2021MNRAS.508.2071A}.

Our study leverages the latest iteration of {\small The Three Hundred} collaboration's suite of cosmological galaxy formation simulations of galaxy clusters, which model a broad range of physical processes - radiative cooling, star formation and supernovae, black hole growth, outflows, and jets - and provide a more realistic treatment of cluster formation than is possible in non-radiative simulations. Nevertheless, we find that a relationship between shock and splashback radii that is consistent with that found in non-radiative simulations, such as those of \citet{2021MNRAS.508.2071A}, who found $r_\mathrm{shock,m}/r_\mathrm{splash} \simeq 1.89$ based on a sample of 65 clusters, compared to our median value of $\simeq 1.91$. Our results also show that both shock and splashback radii correlates with the cluster accretion rate, which is consistent with previous studies. \citet{2021MNRAS.508.1777B} found clusters with high mass fraction of the cluster in substructure, as a proxy for a high accretion rate, tend to have smaller shock and splashback radii, as we show in Figure~\ref{figure:mass_accretion}. 

We note that our results on the relationship between $r_\text{splash}$ and halo mass are broadly consistent with previous work. \citet{2021MNRAS.504.4649O} found that $r_\mathrm{splash}$ decreases with mass for halo masses in the range $10^{13}-10^{15} M_{\odot}$ in the {\small Illustris TNG} simulations, while \citet{Towler.etal.2024} found that $r_\mathrm{splash}$ has a weak negative mass dependence for halos more massive than $10^{14} M_{\odot}$ in the {\small FLAMINGO} simulations. \citet{Towler.etal.2024} also reported a correlation between $r_\mathrm{splash}$ and accretion rate, in agreement with our results. \citet{2021MNRAS.504.4649O} found that $r_\mathrm{splash}$ computed from the gas profile is $\sim10-20\%$ lower than computed using the dark matter profile, while $r_\mathrm{splash}$ computed from the galaxy number density profile (essentially the stellar mass density profile) is similar to that of the dark matter profiles; this is consistent with our results.

Observational limits on the location of accretion shocks in galaxy clusters' outskirts have been recovered by stacking Compton-$y$ maps \autocite{2021MNRAS.508.1777B, 2022MNRAS.514.1645A, 2024MNRAS.527.9378A}. These studies detect an integrated tSZ signal; this is proportional to the line-of-sight integral of the electron pressure, which is related to, but not equal nor proportional to, the gas entropy. \citet{2022MNRAS.514.1645A} locate the accretion shock via a minimum in the logarithmic derivative of the tSZ signal and estimate $r_\mathrm{shock,m}/r_\mathrm{splash} > 2.16\pm0.59$. This is slightly higher than the results suggested by cosmological simulations, but this is not a one-to-one comparison. Future X-ray experiments capable of mapping the outskirts of clusters should allow for a more direct comparison with estimates of the shock radius based on gas entropy \citep[e.g.][]{Simionescu.etal.2021}.   

\section{Conclusions}\label{conclusion}
Using 324 simulated galaxy clusters from {\small The Three Hundred} collaboration, we have investigated the relationship between the shock radius, $r_\text{shock}$, which characterises the boundary between a cluster's gaseous outskirts and accreting gas from the cosmic web, and the splashback radius, $r_\text{splash}$, which characterises the boundary between collisionless material orbiting within the cluster and matter that is infalling for the first time. Depending on our definition, we find the shock radius is larger than splashback radius for most, if not all, clusters. If we stack our clusters and estimate $r_\text{splash}$ and $r_\text{shock}$ from the median radial profiles for dark matter density and gas entropy respectively, we find that the median cluster has $r_\text{shock,p} \simeq 1.38 r_\text{splash} (2.58 R_{200})$, estimated from where $K$ reaches its maximum, and $r_\text{shock,m} \simeq 1.91 r_\text{splash} (3.54 R_{200})$, estimated from when its logarithmic slope is a minimum. If we evaluate $r_\text{splash}$ and $r_\text{shock}$ for each cluster individually, we find that the best-fit linear relation increases as $r_\text{shock} \propto 0.65 r_\text{splash}$, independent of definition, and we observe that $r_\text{shock}/r_\text{splash}$ tends to be larger in clusters that have experienced higher recent mass accretion rates, which is driven primarily by strength of the dependence of $r_\text{splash}$ on the accretion rate rather than any dependence of $r_\text{shock}$. We find that $r_\text{shock}/R_{200}$ and $r_\text{splash}/R_{200}$ anti-correlate with virial mass, $M_{200}$, and recent mass accretion history.

These results are consistent with the results of recent studies \citep[e.g.][]{2021MNRAS.508.2071A} but draw on a larger statistical, mass complete, sample of simulated, run using a state-of-the-art galaxy formation model, and calibrated to reproduce the observed galaxy cluster population, building on the work of \citet{Cui.etal.2022}. While this consistency is to be expected - as previous work has shown \citep[e.g.][]{Power.etal.2020}, the key properties of galaxy cluster outskirts are shaped by the physics of gravitational dynamics and strong hydrodynamic shocks - it is important to verify it. These results also confirm that analytical models that assume the coincidence of $r_\text{shock}$ and $r_\text{splash}$ \autocite[e.g.][]{Patej&Loeb2015} need to be modified, and need to account for mass accretion history and larger scale environment.

Our work has potentially interesting consequences for observational studies of the outskirts of clusters, and efforts to measure empirically the accretion shock. Measurements of $r_\text{splash}$ and phase space caustics using cluster galaxies \citep[e.g.][]{Deason.etal.2021} could offer the means to constrain the recent mass accretion history. This could help to predict the projected radial scale at which we might expect to detect the accretion shock, based on the relationship we have measured in our sample of clusters, which would help to guide measurements of non-thermal emission with radio telescopes \citep[e.g.][]{Vernstrom.etal.2023}, X-ray emission \citep[e.g.][]{Ichikawa.etal.2013,Simionescu.etal.2021,McCall.etal.2024} and the thermal Sunyaev–Zeldovich (tSZ) effect \citep{2024MNRAS.527.9378A}, especially when stacking is required to boost sensitivity. Future work will focus on using mock observables to verify the most reliable methods to recover accurate combined measurements of $r_\text{splash}$ and $r_\text{shock}$.

\begin{acknowledgement}
We thank the anonymous referee for their report. 
This work has been made possible by the {\small The Three Hundred} collaboration\footnote{\url{https://www.the300-project.org}}. The HD simulations (7K and 15K runs) were performed on the MareNostrum Finisterrae3, and Cibeles Supercomputers through The Red Española de Supercomputación grants (AECT-2022-3-0027, AECT-2023-1-0013, AECT-2023-2-0004, AECT-2023-3-0023, AECT-2024-1-0026), on the DIaL3 -- DiRAC Data Intensive service at the University of Leicester through the RAC15 grant: Seedcorn/ACTP317, and on the Niagara supercomputer at the SciNet HPC Consortium. DIaL3 is managed by the University of Leicester Research Computing Service on behalf of the STFC DiRAC HPC Facility (\url{https://www.dirac.ac.uk}). The DiRAC service at Leicester was funded by BEIS, UKRI and STFC capital funding and STFC operations grants. DiRAC is part of the UKRI Digital Research Infrastructure. This work also used the DiRAC Complexity system, operated by the University of Leicester IT Services, which forms part of the STFC DiRAC HPC Facility (\url{https://www.dirac.ac.uk}). This equipment is funded by BIS National E-Infrastructure capital grant ST/K000373/1 and STFC DiRAC Operations grant ST/K0003259/1. DiRAC is part of the National e-Infrastructure. Some of the analysis presented in this work was performed on the OzSTAR national facility at Swinburne University of Technology. The OzSTAR programme receives funding in part from the Astronomy National Collaborative Research Infrastructure Strategy (NCRIS) allocation provided by the Australian Government, and from the Victorian Higher Education State Investment Fund (VHESIF) provided by the Victorian Government. SciNet \citep{Loken_2010} is funded by Innovation, Science and Economic Development Canada; the Digital Research Alliance of Canada; the Ontario Research Fund: Research Excellence; and the University of Toronto.
\end{acknowledgement}

\paragraph{Funding Statement}
MZ is supported by China Scholarship Council. KW and AS acknowledge the support of the Australian Government Research Training Program Fees Offset. AS acknowledges the Bruce and Betty Green Postgraduate Research Scholarship and The University Club of Western Australia Research Travel Scholarship. AS, KW, and CP acknowledge the support of the ARC Centre of Excellence for All Sky Astrophysics in 3 Dimensions (ASTRO 3D), through project number CE170100013. WC is supported by the Atracci\'{o}n de Talento Contract no. 2020-T1/TIC-19882 was granted by the Comunidad de Madrid in Spain, and the science research grants were from the China Manned Space Project. He also thanks the Ministerio de Ciencia e Innovación (Spain) for financial support under Project grant PID2021-122603NB-C21 and HORIZON EUROPE Marie Sklodowska-Curie Actions for supporting the LACEGAL-III project with grant number 101086388. MZ, YL, and XZ acknowledge the support of the National SKA Program of China (Grant Nos. 2022SKA0110200, 2022SKA0110203), the National Natural Science Foundation of China (Grant Nos. 12473001, 11975072, 11835009), and the 111 Project (Grant No. B16009). 
The work presented here emerged out of the annual The300 workshop held at UAM's La Cristalera during the week July 8-12, 2024, partially funded by the 'Ayuda para la Organización de Jornadas Científicas en la UAM en el Marco del Programa Propio de Investigación y con el Apoyo del Consejo Social de la UAM'

\paragraph{Data Availability Statement}
All data used in our analysis are from {\small The Three Hundred} collaboration.

\printendnotes
\printbibliography

\appendix

\section{Sensitivity to Mass Resolution and Galaxy Formation Model}
\label{sec:appendix}
\begin{figure*}[ht!]
\centering
\begin{minipage}{0.48\textwidth}
    \centering
    \includegraphics[width=\linewidth]{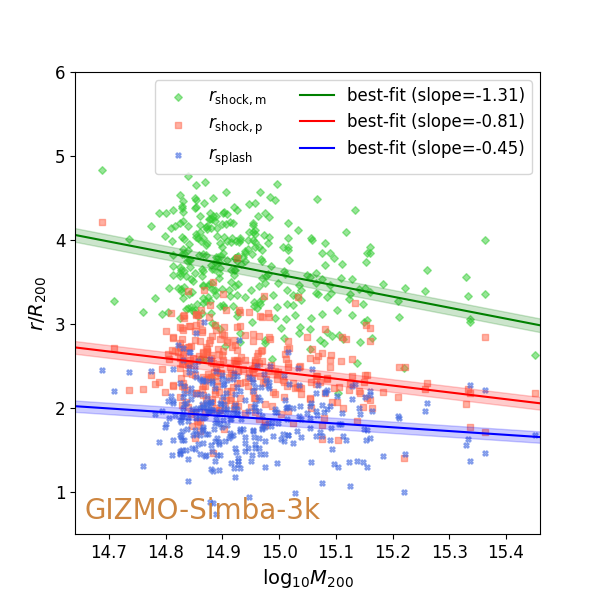}
\end{minipage}
\hfill
\begin{minipage}{0.48\textwidth}
    \centering
    \includegraphics[width=\linewidth]{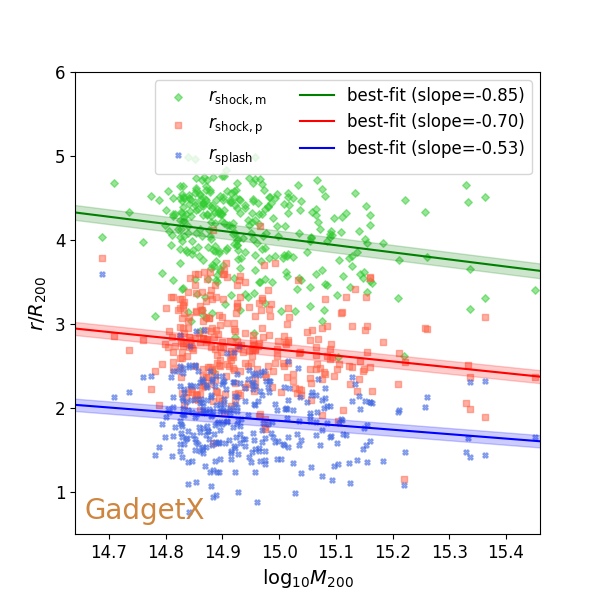}
\end{minipage}


\begin{minipage}{0.48\textwidth}
    \centering
    \includegraphics[width=\linewidth]{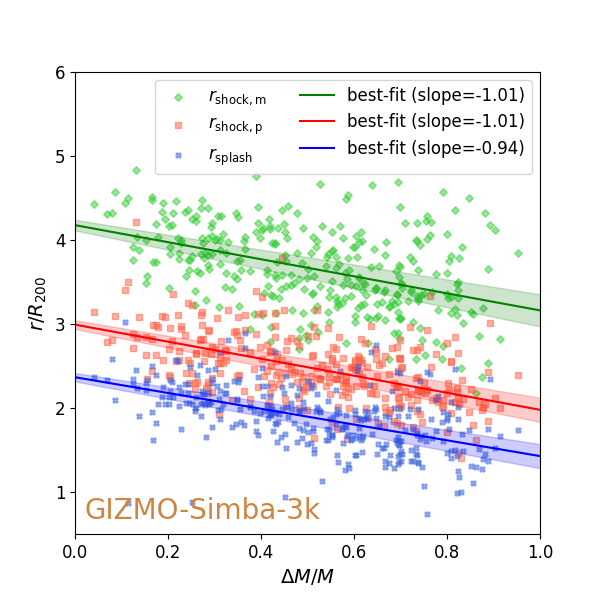}
\end{minipage}
\hfill
\begin{minipage}{0.48\textwidth}
    \centering
    \includegraphics[width=\linewidth]{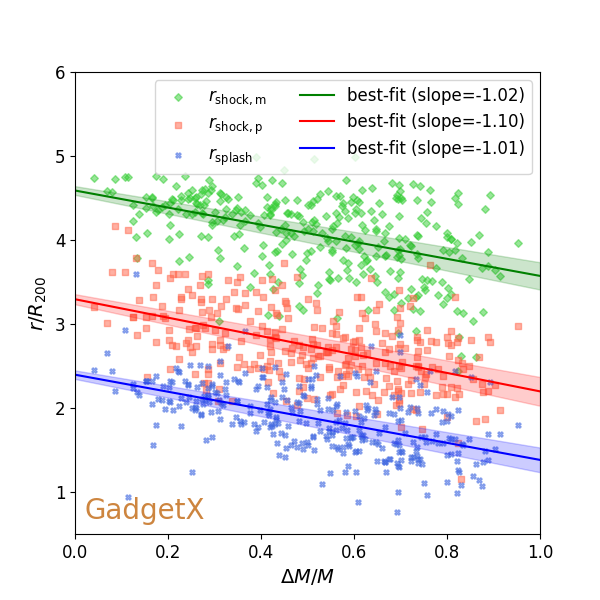}
\end{minipage}
\caption{The relationship between the shock and splashback radii, $r_\text{shock, m}$, $r_\text{shock, p}$ and $r_\text{splash}$ as a function of virial mass, $M_{200}$ and recent mass accretion history $\Delta M/M$ in the {\small GIZMO-Simba-3k} (left two panels) and {\small GadgetX} (right two panels) runs for each of the 324 clusters in our sample. The shaded bands indicate the 1-$\sigma$ variations for each set of points estimated by bootstrapping.}
\label{figure:sensitivity_to_numerical_parameters}
\end{figure*}

We have checked the sensitivity of our results to mass resolution, by comparing measurements for the {\small GIZMO-Simba-7k} shown here and the {\small GIZMO-Simba-3k} runs, and galaxy formation model, by comparing both sets of the {\small GIZMO-Simba} runs to the {\small GadgetX} runs \citep[cf.][]{Cui.etal.2018}. We see similar qualitative trends regardless of mass resolution or galaxy formation model.

The relation between $r_\text{shock,p}$ to $r_\text{splash}$ predicted by {\small GIZMO-Simba-7k}, {\small GIZMO-Simba-3k}, and {\small GadgetX} in the form of   Equation~\ref{eq:shockp_splash} is given by the pairs of coefficients (0.64,1.39), (0.54, 1.45), and (0.58, 1.64) respectively. And the relation between $r_\text{shock,m}$ to $r_\text{splash}$ predicted by {\small GIZMO-Simba-7k}, {\small GIZMO-Simba-3k}, and {\small GadgetX} in the form of   Equation~\ref{eq:shockm_splash} is given by the pairs of coefficients (0.65,2.38), (0.41, 2.88), and (0.47, 3.19) respectively. Similarly, figure \ref{figure:sensitivity_to_numerical_parameters} shows the relationship between shock and splashback radii as a function of $M_{200}$ and $\Delta\,M/M$, for the relations encoded in 
\begin{itemize}
    \item $r_\text{shock,p}/R_{200}$ versus $M_{200}$ (Equation~\ref{eq:shockp_m200}):
(-0.82,15.02), (-0.81,14.59), and (-0.70,13.17)
    \item $r_\text{shock,m}/R_{200}$ versus $M_{200}$ (Equation~\ref{eq:shockm_m200}):
(-1.05,19.22), (-1.31,23.26), and (-0.85,16.73)
    \item $r_\text{splash}/R_{200}$ versus $M_{200}$ (Equation~\ref{eq:splash_m200}):
(-0.56,10.23), (-0.45, 8.59), and (-0.53,9.78)
    \item $r_\text{shock,p}/R_{200}$ versus $\Delta\,M/M$ (Equation~\ref{eq:shockp_deltam}):
(-1.05,3.13), (-1.01, 2.99), and (-1.10,3.29)
    \item $r_\text{shock,m}/R_{200}$ versus $\Delta\,M/M$ (Equation~\ref{eq:shockm_deltam}):
(-1.05,3.13), (-1.01, 4.18), and (-1.02,4.59)
    \item $r_\text{splash}/R_{200}$ versus $\Delta\,M/M$ (Equation~\ref{eq:splash_deltam}):
(-0.98,2.36), (-0.94, 2.37), and (-1.01,2.40)
\end{itemize}
The trends between $r_\text{splash}/R_{200}$, $r_\text{shock,m}/R_{200}$, $r_\text{shock,p}/R_{200}$ and $\Delta\,M/M$ are similar across the different resolutions and galaxy formation models. There are differences in the strength of the anti-correlation between the $R_{200}$ normalised values of $r_\text{splash}$, $r_\text{shock,m}$, $r_\text{shock,p}$, and $M_{200}$ between runs; there is better quantitative agreement between $r_\text{shock,p}/R_{200}$ and $M_{200}$ between the {\small GIZMO-Simba-7k} run and the {\small GIZMO-Simba-3k} run than between the {\small GIZMO-Simba-7k} run and the {\small GadgetX}. This partly reflects the philosophy underpinning the {\small GadgetX} runs, which were calibrated to recover the properties of ICM of observed clusters and partly the improvement in calibration of the  {\small GIZMO-Simba-7k} runs compared to the {\small GIZMO-Simba-3k} runs, which have produced ICM  properties more consistent with observations. This explains the stronger scaling of $r_\text{shock,p}$ with $M_{200}$ and the larger spread in values at a given $M_{200}$ compared to the two other models.

We conclude that, provided care is taken to calibrate runs to recover ICM properties that are consistent with observed clusters, the relations between $r_\text{shock}$, $r_\text{splash}$, $M_{200}$, and $\Delta\,M/M$ are consistent across mass resolution and galaxy formation model.

\end{document}